\documentclass[10pt,twocolumn,superscriptaddress,showpacs,preprintnumbers,amsmath,amssymb,prl,aps]{revtex4}
\usepackage{graphicx}

\begin{document}


\title{Magnetic phase separation and superfluid density measurements in alkali metal-organic solvent intercalated iron selenide superconductor Li$_x$(C$_5$H$_5$N)$_y$Fe$_{2-z}$Se$_2$}

\author{P. K. Biswas}
\email[]{pabitra.biswas@psi.ch}
\affiliation{Laboratory for Muon Spin Spectroscopy, Paul Scherrer Institute, CH-5232 Villigen PSI, Switzerland}

\author{A. Krzton-Maziopa}
\affiliation{Laboratory for Developments and Methods, Paul Scherrer Institute, CH-5232 Villigen PSI, Switzerland}
\affiliation{Faculty of Chemistry, Warsaw University of Technology, Noakowskiego 3, 00-664 Warsaw, Poland}

\author{R. Khasanov}
\affiliation{Laboratory for Muon Spin Spectroscopy, Paul Scherrer Institute, CH-5232 Villigen PSI, Switzerland}

\author{H. Luetkens}
\affiliation{Laboratory for Muon Spin Spectroscopy, Paul Scherrer Institute, CH-5232 Villigen PSI, Switzerland}

\author{E. Pomjakushina}
\affiliation{Laboratory for Developments and Methods, Paul Scherrer Institute, CH-5232 Villigen PSI, Switzerland}

\author{K. Conder}
\affiliation{Laboratory for Developments and Methods, Paul Scherrer Institute, CH-5232 Villigen PSI, Switzerland}

\author{A. Amato}
\affiliation{Laboratory for Muon Spin Spectroscopy, Paul Scherrer Institute, CH-5232 Villigen PSI, Switzerland}

\date{\today}

\begin{abstract}
We report the low-temperature electronic and magnetic properties of the alkali metal-organic solvent intercalated iron selenide superconductor Li$_x$(C$_5$H$_5$N)$_y$Fe$_{2-z}$Se$_2$ using muon-spin-spectroscopy measurements. The zero-field $\mu$SR results indicate that nearly half of the sample is magnetically ordered and there is a spatial phase separation between the superconducting and the magnetic fraction. The transverse-field $\mu$SR results show that the temperature dependence of the penetration depth $\lambda\left(T\right)$ in the mixed state of Li$_x$(C$_5$H$_5$N)$_y$Fe$_{2-z}$Se$_2$ can be explained using a two-gap $s+s$-wave model with gap values of $6.82(92)$ and $0.93(7)$~meV. This implies that the symmetry of the superconducting gap in this system remains unaltered to the parent compound FeSe even after the intercalation with the molecular spacer layer. We obtain $\lambda (0)=485(21)$~nm at $T=0$~K. 
\end{abstract}



\maketitle

Since the discovery of the iron-based superconductor LaFeAsO$_{1-x}$F$_x$ in 2008~\cite{Kamihara}, it has been fascinating to observe how the research evolved during the last four years. In general there are two classes of iron-based superconductors, chalcogenides and pnictides. It was the iron pnictide system which dominated for the first couple of years due to their higher $T_c$ values, with a highest $T_c$ of 56~K for Gd$_{1-x}$Th$_x$FeAsO~\cite{Wang}. Now, the iron chalcogenide system has generated lot of interest to the condensed matter physicists with the development of new superconductors with more and more higher $T_c$ values and a fascinating coexistence and competition with strongly magnetic phases~\cite{Guo,Maziopa1,Shermadini1,Torchetti,Yu,Ma,Kotegawa}. The parent compound of this system is FeSe, superconducting with a $T_c$ of 8.0~K at ambient pressure~\cite{Hsu} and 37~K at 7~GPa~\cite{Margadonna}. The $T_c$ value can be raised as high as 65~K for a single layered FeSe films grown on SrTiO$_3$ substrate~\cite{Wang2,Liu,He}. The substitution of tellurium on the selenium site also increases $T_c$ to maximum of 14.5~K at ambient pressure~\cite{Yeh,Fang} and 23.3~K at 3~GPa~\cite{Gresty}. A further milestone in the evolution of iron chalcogenide superconductors has been achieved by intercalating alkali metals (K, Cs, Rb) between FeSe layers which increases the $T_c$ values above 30~K~\cite{Guo,Maziopa1,Wang1}. These alkali intercalated materials also have a second superconducting phase beyond 12~GPa with a $T_c$ of 48~K~\cite{Sun}. Interest has been redoubled in this class of superconductors with the recent discovery of the enhancement of $T_c$ of FeSe above 40~K by the intercalation of molecular spacer layer~\cite{Ying,Lucas,Scheidt,Maziopa}. It has been claimed that the value of $T_c$ increases with increasing inter-layer distance between the FeSe layers~\cite{Zhang1}. This has stimulated a tremendous research interest to elucidate the superconducting pairing mechanism, and gap symmetry and how they evolve with the increasing inter-layer distances in these intercalated FeSe superconductors.

In this letter, we report detailed investigation of the superfluid density and the symmetry of the superconducting gap in the alkali metal-organic solvent intercalated iron selenide superconductor Li$_x$(C$_5$H$_5$N)$_y$Fe$_{2-z}$Se$_2$ by muon spin rotation/relaxation ($\mu$SR) technique. Our results clearly demonstrate that nearly half of the volume fraction of Li$_x$(C$_5$H$_5$N)$_y$Fe$_{2-z}$Se$_2$ is in a magnetically ordered state. The other half is paramagnetic and becomes fully superconducting below $T_c$. The observed $\lambda^{-2}\left(T\right)$ is found to be well described by a similar two-gap $s+s$-wave model, as seen in the parent compound FeSe$_{0.85}$~\cite{Khasanov}. Comparing these results with another newly discovered intercalated iron selenide superconductor Li$_{0.6}$(NH$_2$)$_{0.2}$(NH$_3$)$_{0.8}$Fe$_2$Se$_2$~\cite{Lucas}, we suggest that the 2~dimentional (D) FeSe layer generate superconductivity in this class of materials and the value of $T_c$ depends only on the superfluid density within this FeSe layer which does not change with the increased interlayer distance.

$\mu$SR measurements were performed using the general purpose surface (GPS) muon instrument located on the $\pi$M3.2 beamline of the Swiss Muon Source at the Paul Scherrer Institute, Villigen, Switzerland. For details about this technique, see Refs.~\cite{Dalmas,Amato,Hillier,Blundell,Sonier,Brandt}. Data were collected both with zero field (ZF) and transverse field (TF) modes. In the TF mode, an external magnetic field ($H=30$~Oe) was applied perpendicular to the initial direction of the muon spin polarization. The magnetic field was applied above the superconducting transition temperature and the sample then cooled to base temperature, i.e. so-called field-cooled (FC) procedure. A continuous-flow ${}^4$He cryostat was used to collect the data between 1.6 to 250~K. The number of positron events were 18 million for each data point.

A polycrystalline sample of Li$_x$(C$_5$H$_5$N)$_y$Fe$_{2-z}$Se$_2$ ($x\approx1.0$, $y\approx0.2$ and $z\approx2.0$) was prepared via room temperature intercalation into the iron selenide matrix in pyridine solution of the Li metal as described in Ref.~\cite{Maziopa}. A circular disk shaped pellet of the powder sample of Li$_x$(C$_5$H$_5$N)$_y$Fe$_{2-z}$Se$_2$ with 8~mm diameter and 2~mm thickness was mounted on a sample holder. The pellet was covered with a thin layer of polymer and kept in an inert atmosphere to inhibit air decomposition.

\begin{figure}[tb!]
\begin{center}
\includegraphics[width=0.7\columnwidth]{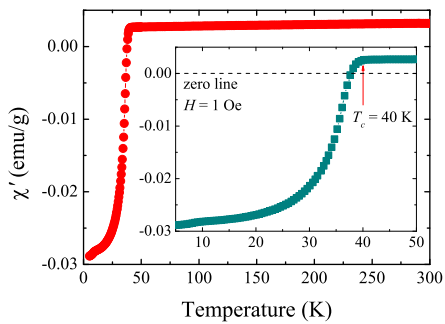}
\caption{\label{Figure1Biswas} (Color online) Temperature dependence of the ac-magnetic susceptibility (real part only) of Li$_x$(C$_5$H$_5$N)$_y$Fe$_{2-z}$Se$_2$ measured using zero-field-cooled mode under a magnetic field of $H_{ac}=1$~Oe. The inset shows the diamagnetic transition in Li$_x$(C$_5$H$_5$N)$_y$Fe$_{2-z}$Se$_2$ at 40(1)~K.}
\end{center}
\end{figure}

Temperature dependence of the ac-magnetic susceptibility of Li$_x$(C$_5$H$_5$N)$_y$Fe$_{2-z}$Se$_2$ was performed in a Quantum Design PPMS magnetometer under a magnetic field of $H_{ac}=1$~Oe, using zero-field-cooled (ZFC) protocol. Figure~\ref{Figure1Biswas} shows the real part of the ZFC ac susceptibility of Li$_x$(C$_5$H$_5$N)$_y$Fe$_{2-z}$Se$_2$, measured at temperature ranging from 5~K to 300~K. The inset of Fig.~\ref{Figure1Biswas} shows a diamagnetic transition in the susceptibility data with a $T_c$ onset at 40(1)~K. A rather large positive signal in the normal state susceptibility data shows the presence of a sizeable amount of ferromagnetic impurity in the sample. This may be due to the presence of some small clusters of Fe ions and other Fe-based magnetic phases such as $\alpha$-FeSe, Fe$_7$Se$_8$, Fe$_3$O$_4$ in the sample. Comparing the normal state susceptibility value with the saturation magnetization of elemental Fe, Fe$_7$Se$_8$ and Fe$_3$O$_4$ we estimate that the amount of this magnetic impurity is not more than 1~\% of the sample mass. A similar scenario has also been observed in many other Fe-based superconductors~\cite{Khasanov,Zhang,Patel,McQueen,Rongwei,Maziopa,Lucas}.

\begin{figure}[tb!]
\begin{center}\includegraphics[width=0.7\columnwidth]{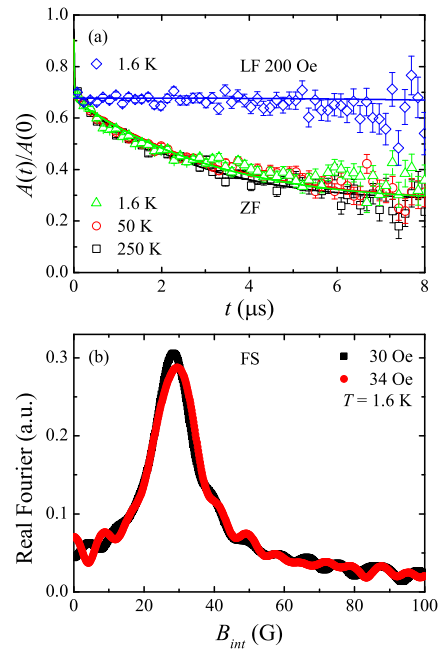}
\caption{\label{Figure2Biswas} (Color online) (a) ZF-$\mu$SR time spectra collected at 1.6~K, 50~K and 250~K for Li$_x$(C$_5$H$_5$N)$_y$Fe$_{2-z}$Se$_2$. LF-$\mu$SR time spectra collected at 1.6~K and 200~Oe. The solid lines are the fits to the data using the equation described in the text. (b) Internal field distribution, $B_{int}$ in the vortex state of Li$_x$(C$_5$H$_5$N)$_y$Fe$_{2-z}$Se$_2$ for the two different applied fields of 30 and 34~Oe.}
\end{center}
\end{figure}

In order to search for any magnetic anomalies (static or fluctuating) in Li$_x$(C$_5$H$_5$N)$_y$Fe$_{2-z}$Se$_2$, we have performed a ZF-$\mu$SR study. Figure~\ref{Figure2Biswas}~(a) shows the ZF-$\mu$SR signals collected at 1.6~K, 50~K and 250~K for Li$_x$(C$_5$H$_5$N)$_y$Fe$_{2-z}$Se$_2$. The signals are almost identical in nature and there is no precessional signal and no big change in the observed relaxation rate between data collected above and below $T_c$. However, we have observed a large drop of the asymmetry value of the ZF $\mu$SR signal, occurring within the first 100~ns, over the entire temperature range. This suggests that some portion of our sample is purely magnetic in nature. To calculate the volume fraction of the magnetic part, we have fitted the following equation to the ZF data:

\begin{equation}\label{ZF_Fit}
A^{ZF}(t)=v_mA^{ZF}_m(t)+(1-v_m)A^{ZF}_{nm}(t)
\end{equation}
where
\begin{subequations}\label{ZF_Fit_1}
\begin{align}
A^{ZF}_m(t)&=A^{ZF}(0)\left[\frac{2}{3}\exp(-\lambda_Tt)+\frac{1}{3}\exp(-\lambda_Lt)\right]\\
A^{ZF}_{nm}(t)&=A^{ZF}(0)\left[\frac{1}{3}+\frac{2}{3}(1-\sigma^{2}t^{2}-\Lambda{t})
\exp(-\frac{\sigma^{2}t^{2}}{2}-\Lambda{t})\right]
\end{align}
\end{subequations}
Here \textit{m} and \textit{nm} denote the magnetic and nonmagnetic components, $A^{ZF}(0)$ is the initial asymmetry, $v_m$ is the magnetic volume fraction, and $\lambda_T$, $\lambda_L$ are the transverse and longitudinal components of the muon decay for the magnetic fraction of the sample, respectively. $A^{ZF}_{nm}(t)$ is the combination of Lorentzian and Gaussian
Kubo-Toyabe (LGKT) relaxation function~\cite{Kubo,Hayano} for the nonmagnetic fraction of the sample with parameters $\sigma$ and $\Lambda$, the muon depolarization rates arise due to the concentrated nuclear dipole moments and randomly orientated diluted local electronic moments, respectively. The fits yield the parameters shown in Table~\ref{table_of_fits}. Parameters obtained from the fitting show that nearly half of the sample is antiferromagnetically (AFM) ordered. This is also consistent with other Fe-based superconductors~\cite{Shermadini}. The low relaxation rate $\Lambda$ most probably stem from the diluted ferromagnetic Fe impurity clusters. This hypothesis is supported by the fact that it is possible to decouple this slow relaxation with the application of a small LF of 200~Oe (see Fig.~\ref{Figure2Biswas}~(a)). This also proves that the dipole field at the muon site is static on the time scale of the $\mu$SR window. The magnetic volume fraction is practically temperature independent between 1.6 and 250~K. This strongly resembles the situation of phase separation in the alkali-metals intercalated FeSe~\cite{Shermadini,Shermadini1}. Our data show that there is a clear phase separation between the magnetic and the nonmagnetic fractions and hence the magnetic fraction is not affecting the properties of the nonmagnetic part of the sample.

\begin{table}
\caption{Parameters extracted from the fits using Eqs.~\ref{ZF_Fit}~and~\ref{ZF_Fit_1} to the zero-field-$\mu$SR data collected above and below $T_{c}$ for Li$_x$(C$_5$H$_5$N)$_y$Fe$_{2-z}$Se$_2$.}
\label{table_of_fits}
\begin{center}
\begin{tabular}[t]{lll}\hline\hline
$T$ (K)~~~~~~&$v_m$ (\%)~~~~~~~~~~~~&$\Lambda$\\\hline
1.6 & 49.05$\pm$0.6 & 0.18$\pm$0.01\\
50 & 50.8$\pm$0.6 & 0.19$\pm$0.01\\
250 & 49.1$\pm$0.6 & 0.20$\pm$0.01\\\hline\hline
\end{tabular}
\par\medskip\footnotesize
\end{center}
\end{table}

We have also performed field shift (FS) measurements on Li$_x$(C$_5$H$_5$N)$_y$Fe$_{2-z}$Se$_2$. Figure~\ref{Figure2Biswas}~(b) shows the internal field distribution, $B_{int}$ in the vortex state of Li$_x$(C$_5$H$_5$N)$_y$Fe$_{2-z}$Se$_2$ for the two different applied fields of 30 and 34~Oe. FS measurements show that an application of additional 4~Oe magnetic field below $T_c$ does not shift the internal field distribution by a significant amount ($\approx$~0.5~G only) due to vortex pinning. This shows that the paramagnetic volume fraction of the sample becomes fully superconducting below $T_c$.

\begin{figure}[tb!]
\begin{center}
\includegraphics[width=0.7\columnwidth]{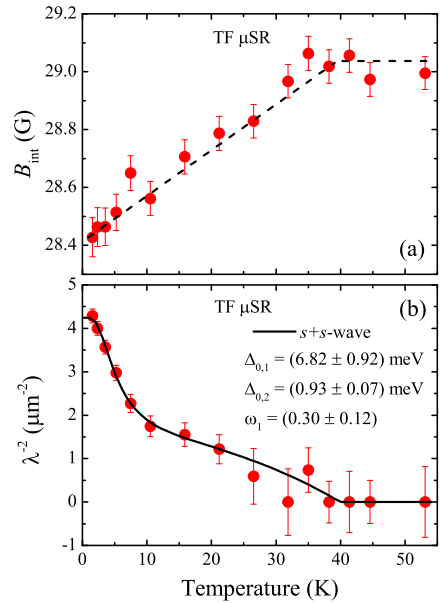}
\caption{\label{Figure3Biswas} (Color online) (a) Temperature dependence of the internal magnetic field at the muon sites in Li$_x$(C$_5$H$_5$N)$_y$Fe$_{2-z}$Se$_2$. The dashed line is simply a guide to the eye. (b) Temperature dependence of $\lambda^{-2}$ for Li$_x$(C$_5$H$_5$N)$_y$Fe$_{2-z}$Se$_2$. The curve (black line) is a fit to the data using two $s$-wave components, each with an isotropic gap.}
\end{center}
\end{figure}

To reveal the superconducting properties on Li$_x$(C$_5$H$_5$N)$_y$Fe$_{2-z}$Se$_2$, we have performed a TF-$\mu$SR study. It was very difficult to extract the superconducting properties of Li$_x$(C$_5$H$_5$N)$_y$Fe$_{2-z}$Se$_2$ using a higher magnetic field due to higher relaxation from the diluted ferromagnetic impurities present in the sample. Therefore, we have used a small magnetic field of 30~Oe for this TF-$\mu$SR study. It is noteworthy to mention that the lower critical field of all the iron chalcogenide superconductors are very low and in our case, 30~Oe was high enough to drive Li$_x$(C$_5$H$_5$N)$_y$Fe$_{2-z}$Se$_2$ into a well defined vortex state. Figure~\ref{Figure3Biswas} (a) shows the temperature dependence of the internal magnetic field at the muon sites in Li$_x$(C$_5$H$_5$N)$_y$Fe$_{2-z}$Se$_2$. The dashed line is drawn as a guide for the eye. We observed an expected diamagnetic shift of the internal magnetic field experienced by the muons below $T_c$. 

To extract the superfluid density of Li$_x$(C$_5$H$_5$N)$_y$Fe$_{2-z}$Se$_2$, we have fitted the TF muon time spectra using an oscillatory decaying Gaussian function,

\begin{eqnarray}
\label{Depolarization_Fit}
A^{TF}(t)=A^{TF}(0)\exp\left(-\sigma^{2}t^{2}\right/2)\cos\left(\gamma_\mu B_{int}t +\phi\right)  \nonumber \\
+A^{TF}_{bgd}(0)\cos\left(\gamma_\mu B_{bgd}t +\phi\right),~
\end{eqnarray}
where $A^{TF}(0)$ is the initial asymmetry, $\gamma_{\mu}/2\pi=135.5$~MHz/T is the muon gyromagnetic ratio~\cite{Sonier}, $B_{int}$ and $B_{bgd}$ are the internal and background magnetic field at the muon sites, $\phi$ is the initial phase offset of the muon precession signal, respectively and $\sigma$ is the Gaussian muon spin relaxation rate. It can be written as $\sigma=\left(\sigma^{2}_{sc} + \sigma^{2}_{nm}\right)^{\frac{1}{2}}$, where $\sigma_{sc}$ is the superconducting contribution to the relaxation rate due to the field variation across the flux line lattice and $\sigma_{nm}$ is the nuclear magnetic dipolar contribution which is assumed to be constant over the temperature range of the study.

In a superconductor with a large upper critical field and a hexagonal Abrikosov vortex lattice, the Gaussian muon-spin depolarization rate $\sigma_{sc}$ is related to the penetration depth $\lambda$ by the expression

\begin{equation}
\frac{\sigma_{sc}^{2}\left(T\right)}{\gamma_\mu^2}=0.00371\frac{\Phi_0^2}{\lambda^{4}\left(T\right)},
\end{equation}
where $\Phi_0=2.068\times10^{-15}$~Wb is the flux quantum~\cite{Brandt}. Figure~\ref{Figure3Biswas} (b) shows the temperature dependence of $\lambda^{-2}$ for Li$_x$(C$_5$H$_5$N)$_y$Fe$_{2-z}$Se$_2$.

The temperature dependence of the penetration depth of Li$_x$(C$_5$H$_5$N)$_y$Fe$_{2-z}$Se$_2$ can be fitted using a two-gap $s+s$-wave model~\cite{Carrington, Padamsee},

\begin{equation}
\label{two_gap}
\frac{\lambda^{-2}\left(T\right)}{\lambda^{-2}\left(0\right)}=\omega\frac{\lambda^{-2}\left(T, \Delta_{1}\right)}{\lambda^{-2}\left(0,\Delta_{1}\right)}+(1-\omega)\frac{\lambda^{-2}\left(T, \Delta_{2}\right)}{\lambda^{-2}\left(0,\Delta_{2}\right)},
\end{equation}
where $\lambda\left(0\right)$ is the value of the penetration depth at $T=0$~K, $\Delta_{i}$ is the value of the $i$-th ($i=1$ or 2) superconducting gap at $T=0$~K and $\omega$ is the weighting factor of the first gap~\cite{gapnote}.

Each component in Eq.~\ref{two_gap} can be expressed within the local London approximation~\cite{Tinkham,Prozorov} as

\begin{equation}
\frac{\lambda^{-2}\left(T, \Delta_{i}\right)}{\lambda^{-2}\left(0, \Delta_{i}\right)}=1+2\int^{\infty}_{\Delta_{i}}\left(\frac{\partial f}{\partial E}\right)\frac{ EdE}{\sqrt{E^2-\Delta_{i}\left(T\right)^2}},
\end{equation}
where $f=\left[1+\exp\left(E/k_BT\right)\right]^{-1}$ is the Fermi function, and $\Delta_i\left(T\right)=\Delta_{i}\delta\left(T/T_c\right)$. The temperature dependence of the gap is approximated by the expression $\delta\left(T/T_c\right)=\tanh\left\{1.82\left[1.018\left(T_c/T-1\right)\right]^{0.51}\right\}$~\cite{Carrington}.

The curve shown in Fig.~\ref{Figure3Biswas}~(b) is a fit of the two-gap $s+s$-wave model to the data. The fit yields $\Delta_{1}=6.82(92)$~meV and $\Delta_{2}=0.93(7)$~meV with $\omega=0.30(12)$. The ratio of the larger to the smaller gap, $\Delta_{1}/\Delta_{2}\sim7.3(2)$ found in Li$_x$(C$_5$H$_5$N)$_y$Fe$_{2-z}$Se$_2$ is larger than the corresponding value (4.2) seen in FeSe$_{0.85}$ but is consistent with the value (7.7) for the lithium amide and ammonia intercalated FeSe determined by measuring the magnetic penetration depth using the $\mu$SR technique~\cite{Khasanov,Lucas}. The large gap to the $T_c$ ratio is $2\Delta_{1}/k_{B}T_c=3.96(54)$ and the small gap to the $T_c$ ratio is $2\Delta_{2}/k_{B}T_c=0.57(4)$. These suggest that the large gap lies above the strong-coupling limit whereas the small gap are in the weak-coupling limit. Table~\ref{table_of_gapratios} summarises the superconducting gap to the $T_c$ ratios for different iron chalcogenide superconductors by means of $\mu$SR study.

\begin{table}
\caption{Superconducting gap to the $T_c$ ratios for different iron chalcogenide superconductors by means of $\mu$SR study.}
\label{table_of_gapratios}
\begin{center}
\begin{tabular}[t]{llll}\hline\hline
Compounds &$2\Delta_{1}/k_{B}T_c$ &$2\Delta_{2}/k_{B}T_c$ &Refs.\\\hline
FeSe$_{0.85}$ & 4.49(6) & 1.07(2) & \cite{Khasanov}\\
FeTe$_{0.5}$Se$_{0.5}$ & 4.19(16) & 1.40(9) & \cite{Biswas}\\
K$_{0.74}$Fe$_{1.66}$Se$_{2}$ & 4.7(2) &  & \cite{Shermadini}\\
Rb$_{0.77}$Fe$_{1.61}$Se$_{2}$ & 5.5(2) &  & \cite{Shermadini}\\
Li$_{0.6}$(NH$_2$)$_{0.2}$(NH$_3$)$_{0.8}$Fe$_2$Se$_2$ & 5.27(21) & 0.69(2) & \cite{Lucas}\\
Li$_x$(C$_5$H$_5$N)$_y$Fe$_{2-z}$Se$_2$ & 3.96(54) & 0.57(4) & current\\\hline\hline
\end{tabular}
\par\medskip\footnotesize
\end{center}
\end{table}


The value of the penetration depth at $T=0$~K is found to be $\lambda\left(0\right)=485(21)$~nm for Li$_x$(C$_5$H$_5$N)$_y$Fe$_{2-z}$Se$_2$. For an anisotropic polycrystalline samples, $\lambda$ is related to $\lambda_{ab}$ by $\lambda=3^{\frac{1}{4}}\lambda_{ab}$~\cite{Fesenko}. With this assumption, we found $\lambda_{ab}\left(0\right)=369(16)$~nm for Li$_x$(C$_5$H$_5$N)$_y$Fe$_{2-z}$Se$_2$. These values are comparable with those obtained by Khasanov, \textit{et al.}~\cite{Khasanov} for FeSe$_{0.85}$ but 50~$\%$ higher than the value obtained from Li$_{0.6}$(NH$_2$)$_{0.2}$(NH$_3$)$_{0.8}$Fe$_2$Se$_2$ by Lucas, \textit{et al.}~\cite{Lucas} using $\mu$SR technique. This discrepency may be due to the fact that the interlayer spacing between the FeSe layers in our compound (lattice parameter, $c=23.09648$~\AA) is 40~$\%$ higher than Li$_{0.6}$(NH$_2$)$_{0.2}$(NH$_3$)$_{0.8}$Fe$_2$Se$_2$ ($c=16.5266$~\AA). Recent $\mu$SR studies on bismuth-based high-$T_c$ cuprate superconductors by Baker \textit{et al.}~\cite{Baker} suggest that the penetration depth in this class of materials increases with increasing layer separation between the CuO$_2$ layers with a relation $1/d\propto1/\lambda^2_{ab}$, where $d$ is the interlayer distance. It is also found that the $T_c$ value depends only on the 2D superfluid density but not on the bulk superfluid density. For Li$_x$(C$_5$H$_5$N)$_y$Fe$_{2-z}$Se$_2$, we obtain $d/\lambda^2_{ab}=1.84(14)\times10^{4}$~m$^{-1}$, which is consistent with the value $2.6\times10^{4}$~m$^{-1}$, estimated for Li$_{0.6}$(NH$_2$)$_{0.2}$(NH$_3$)$_{0.8}$Fe$_2$Se$_2$. The nearly constant ratio of $d/\lambda^2_{ab}$ for the two intercalated FeSe materials point to a noteworthy similarity to the high-$T_c$ cuprate superconductors. Like the CuO$_2$ layer in the high-$T_c$ cuprates, it is the 2D FeSe layer which generates superconductivity in this system and with an increased interlayer distance, the superfluid density within the layer does not change.

In summary, $\mu$SR measurements have been performed on the alkali metal-organic solvent intercalated iron selenide superconductor Li$_x$(C$_5$H$_5$N)$_y$Fe$_{2-z}$Se$_2$ with a $T_c$ of 40(1)~K. The superfluid density of this compound is found to be very low compare to other iron-chalcogenide superconductors probably due to its 2D nature. Nearly $50~\%$ (volume fraction) of the sample is magnetically ordered. A spatial phase separation has been observed between the superconducting and the magnetic fraction of the sample. The temperature dependence of the magnetic penetration depth is found to be compatible with a two-gap $s+s$-wave model with gap values $\Delta_{1}=6.82(92)$~meV and $\Delta_{2}=0.93(7)$~meV. We obtain $\lambda (0)=485(21)$ nm at $T=0$~K. Further studies are in progress to explore the magnetic properties and if there is any microscopic coexistence of superconductivity and magnetism in the magnetic volume of this material.

The $\mu$SR experiments were performed at the Swiss Muon Source, Paul Scherrer Institut, Villigen, Switzerland. A.K-M gratefully acknowledges the financial support from Institute of Physics University of Z$\ddot{\rm{u}}$rich.

\bibliography{Biswas}

\begin{thebibliography}{46}
\bibitem[{gap()}]{gapnote}
\bibinfo{note}{Note that the $\mu$SR technique can not determine the phase of
  the superconducting order parameter and therefore can not distinguish between
  $s^+$ and $s^-$-wave gaps. For the two-gap $s+s$-wave model, we have simply
  added two single $s$-wave gap equations along with a weighting factor.}
\end{thebibliography}

\end{document}